# Crystal Growth of Spin-Frustrated Ba$_4$Nb$_{0.8}$Ir$_{3.2}$O$_{12}$: A Possible Spin Liquid Material

*Gohil S. Thakur* [1]*, *Sumanta Chattopadhyay* [2], *Thomas Doert* [3], *T. Herrmannsdörfer* [2] and *Claudia Felser* [1]

[1] Max Planck Institute for Chemical Physics of Solids, 01187 Dresden, Germany

[2] Dresden High Magnetic Field Laboratory (HLD-EMFL), Helmholtz-Zentrum Dresden-Rossendorf, 01328 Dresden, Germany

[3] Faculty of Chemistry and Food Chemistry, Technical University, 01069, Dresden, Germany

**Abstract**: Polycrystalline Ba$_4$NbIr$_3$O$_{12}$ has recently been shown to be a promising spin liquid candidate. We report an easy and reliable method to grow millimeter-sized single crystals of this trimer–based spin liquid candidate material with the actual stoichiometry of Ba$_4$Nb$_{0.8}$Ir$_{3.2}$O$_{12}$. The growth of large crystals is achieved using BaCl$_2$ as flux. The crystals show hexagonal plate-like habit with edges up to 3 mm in length. The structure is confirmed by single-crystal X-ray diffraction and is found to be the same as of previously reported phase Ba$_{12}$Nb$_{2.4}$Ir$_{9.6}$O$_{36}$ [Ba$_4$Nb$_{0.8}$Ir$_{3.2}$O$_{12}$], indeed with a mixed occupancy of Nb/Ir at 3*a* site. The magnetic and calorimetric study on the individual single crystals confirms the possibility of a spin liquid state consistent with a recent report on a polycrystalline sample.

**Introduction**

Quantum spin liquid (QSL) materials have been a topic of intense research very recently owing to their very rich and interesting magnetic properties like long-range entanglement and fractional quantum excitations without any spontaneous symmetry breaking of the crystal lattice or spins. In simple terms, such materials are characterized by fluctuating spins entangle over long distances without showing magnetic order in the zero-temperature limit. These materials are rare and in fact, the spin liquid state has only been realized experimentally in a handful of compounds mostly with triangular lattice structures.[1–7] All of these materials possess various degrees of spin frustration. Very recently an Ir-based oxide containing $Ir_3O_{12}$ trimers has been proposed as a candidate material based on the magnetic and calorimetric studies on a polycrystalline sample.[8] Magnetic and calorimetric measurements on this sample suggested spin frustration and no magnetic ordering. Although this compound had been first reported as microcrystalline material with crystals only suitable for diffraction studies, no other properties were reported.[9] High-quality single crystals are thus essential to confirm the spin liquid state by various thermodynamic and microscopic measurement techniques such as specific heat, dynamic susceptibility, neutron scattering, etc at low temperatures. Single crystals are also required to ascertain that there are no other underlying magnetic impurity phases, which may contribute to paramagnetism and give rise to other artifacts in low-temperature measurements. Also, it is important to eliminate the possibility of a disorder due to non-stoichiometry or unintentional doping during the sample growth, because in frustrated magnets only a tiny disorder can destroy the spin-liquid state and lead to a spin glass state.

In this work, we report the growth of large millimeter-sized single crystals of the title compound and corroborate its properties with the reported work on the polycrystalline sample. The structure is found to be the same as reported earlier by Müller–Bushbaum, in fact with a significant disorder at the Nb site with Ir occupying as much as 20% of the Nb sites.[9] This disorder was overlooked in the recent study reported by Nguyen *et al*.[8] However, despite the disorder, the magnetic and specific heat data still strongly suggest a spin liquid state might exist in this compound.

**Experimental:**

*Synthesis.* Crystal growth was carried out following two methods, both using $BaCl_2$ as flux. In first, a phase pure polycrystalline material was synthesized by heating an appropriate mixture of $BaO_2$, $Nb_2O_5$ and $IrO_2$ with a target composition of $Ba_4NbIr_3O_{12}$ in an evacuated quartz tube at 1100°C for 48 h. The obtained polycrystalline sample was then mixed with an excess of $BaCl_2·2H_2O$, placed in an alumina crucible, heated to 1100°C for 24 h, and slowly cooled to 950°C at various rates ($x$ °C/h) after which the furnace was allowed to cool naturally. Single crystals could also be grown by directly heating $BaCO_3$, $IrO_2$, and $Nb_2O_5$ with an excess of $BaCl_2·2H_2O$ under the same heating conditions. The crystalline product was obtained after dissolving the flux in distilled water and then sonicating in ethanol to remove any polycrystalline matrix or flux.

*Characterization*: Phase purity and identification of the obtained samples were verified by the powder X-ray diffraction (XRD) technique. The powder patterns were collected in the 2θ range from 3.5 to 100° at room temperature on a HUBER G670 imaging plate Guinier camera with Cu-K$\alpha$1 radiation (λ = 1.5406 Å). Rietveld fitting was performed with the TOPAS-4.2.0.2 (AXS) program.[10] The structure was solved by the Charge-Flipping algorithm[11] and refined with Jana2006[12] based on single-crystal X-ray data. Crystallographic data have been deposited with Fachinformationszentrum Karlsruhe, D-76344 Eggenstein-Leopoldshafen (Germany) and can be obtained on quoting the depository number CSD 1983872.

Compositional analyses of the single crystals from different batches of samples were carried out using SEM-EDS and wet chemical analysis (ICP) techniques.

The crystallographic orientation of the crystal was determined from the Laue diffraction patterns. The plane of the crystal was found to be the crystallographic *ab*–plane.

*Physical property measurements*: DC magnetic susceptibility measurements in several fields ($\mu_0H$) and temperatures ($T$) were performed in a SQUID magnetometer (Quantum Design). The magnetization of a single crystal was measured along two different crystallographic orientations with the field ranging from 0.05 to 3.5 T applied parallel and perpendicular to the $c$–axis ($H\perp ab$ and $H\perp c$) and in the temperature between 2 and 320 K in field-cooled protocols (FC)). Field-dependent-magnetization was measured at 2 K in the field range of ± 7 T.

Specific heat ($C_p$) in the temperature range of 2–20 K was measured using the relaxation method. Electrical resistivity ($\rho$) was measured in the $ab$–plane in the temperature range of 90–400 K by a four-probe method in a Quantum Design PPMS instrument.

**Results and discussion:**

Table 1: Crystal growth conditions for $Ba_4Nb_{0.8}Ir_{3.2}O_{12}$.

| Sample | starting mixture | $BaCl_2\cdot 2H_2O$ (moles) | cooling rate $x$ (°C/h) | crystal size | Phase purity |
|--------|------------------|------------------|-------------------|--------------|--------------|
| A | $BaCO_3$, $Nb_2O_5$, $IrO_2$ | 20 | 1.5 | < 0.4 mm | 10% Ir |
| B | $BaCO_3$, $Nb_2O_5$, $IrO_2$ | 25 | 1.25 | < 0.5 mm | 1% Ir, 5% $Ba_5Nb_3O_{15}$ |
| C | polycrystalline $Ba_4Nb_{0.8}Ir_{3.2}O_{12}$ | 20 | 1.25 | < 0.5 mm | 2% Ir |
| D | $BaCO_3$, $Nb_2O_5$, $IrO_2$ | 25 | 0.85 | ~1-2 mm | 1% Ir |
| E | polycrystalline $Ba_4Nb_{0.8}Ir_{3.2}O_{12}$ | 35 | 0.85 | ~ 3-5 mm | Single phase |

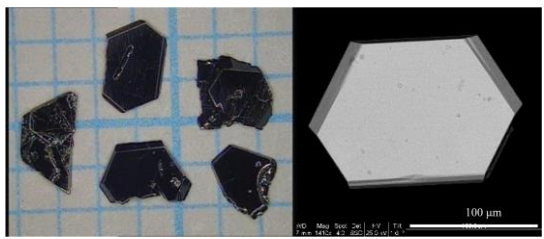

Figure 1. Optical image (left) of $Ba_4Nb_{0.8}Ir_{3.2}O_{12}$ crystals on a millimeter grid and electron microscopic image (right) of a smaller crystal.

Single crystals of $Ba_4Nb_{0.8}Ir_{3.2}O_{12}$ were synthesized by solid-state reaction of starting materials in air employing $BaCl_2$ as flux. The crystals appear as black hexagonal plates with reflective surfaces. A picture of a few representative crystals is shown in figure 1. Crystal quality and size are influenced by both, the choice of the starting materials and the cooling rate. A summary of products obtained using different starting materials and cooling rates is presented in table 1. When the reaction is carried out starting with a mixture of $BaCO_3$, $IrO_2$, and $Nb_2O_5$, the crystal size is generally small and Ir metal is found on the edges or surface of the crystals. Using a polycrystalline powder as charge, this problem is largely eliminated and crystals of larger size are obtained. Larger crystals also tend to grow at a slower cooling rate in combination with the use of polycrystalline material. The quality of crystalline products obtained in each batch was adjudged from the powder X-ray (Figure S1 in SI) and Laue diffraction patterns (Figure S2 in SI). Sharp diffraction spots in Laue patterns confirm that the crystals are of high quality. The flat surface of the crystals is found to be the crystallographic *ab*-plane.

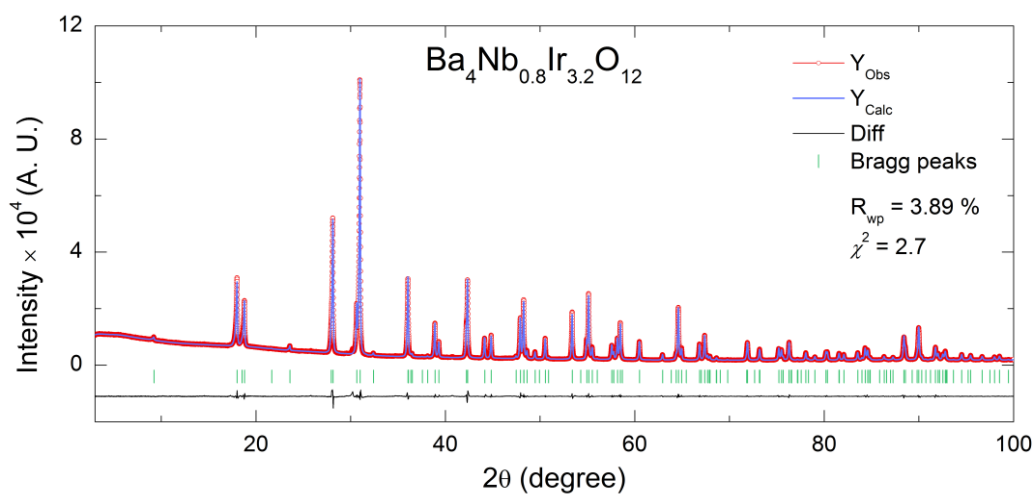

Figure 2. Powder diffraction data (red) of crushed crystals of $Ba_4Nb_{0.8}Ir_{3.2}O_{12}$ and its Rietveld fit (blue). The black line is the difference between the observed and the calculated intensities. Green vertical bars mark the positions of the allowed Bragg reflections.

Figure 2 shows the powder XRD pattern of the crushed single crystals obtained by using polycrystalline powder as the starting material. The formation of a single-phase sample was thus confirmed. A visual inspection of the crystals under an optical microscope showed a clean surface free from any polycrystalline matrix. Semi-quantitative analysis by SEM−EDX shows the presence of all the metals in the average ratio Ba/Ir/Nb = 4.4/3.3/1, confirming the identity of the products. Ir/Nb ratio of greater than 3 indicates Nb might be under-occupied. A table of metal ratio obtained from EDX is shown in Table S1 in SI.

Rietveld refinement of the powder diffraction data was carried out to confirm the structure, results of which are presented in figure 2. All Wyckoff positions used initially were taken from the data reported by Nguyen et al.[8] During the initial Rietveld refined cycle, the occupancies of all the atoms were set free which yielded values very close to unity except for the Nb. An unconstrained refinement of Nb occupancy always yielded a value greater than unity (1.22 to 1.25) which indicated an excess of electron density at the Nb site. Considering similarities in valence states, coordination and ionic radii of Ir and Nb, Ir was allowed to share the Nb site, which led to the occupation factor for Nb, and Ir settling very close to 0.8 and 0.2 respectively. This is a strong indication of a mixed site occupancy at the Nb site. Focusing the structure refinement data reported in,[8] the thermal parameter reported for Nb was found to be very much smaller than for Ba and Ir despite its smaller atomic number, which clearly indicates a higher experimental electron density at Nb site. In order to further confirm the precise stoichiometry of the compound, a structure analysis on a single crystal was performed. The results of the single-crystal analysis are presented in table 2 and the crystal data along with the anisotropic temperature factors are presented in tables S2 and S3, respectively. The structure refinement yielded exactly the same structure as reported earlier by Müller–Buschbaum et al[9]. More importantly, our structural analysis confirmed that there is indeed a mixed occupation of Nb/Ir at the 3$a$ sites, which was overlooked in the powder refinement reported in ref. 8. Thus, the correct stoichiometry of the compound is $Ba_4Nb_{0.8}Ir_{3.2}O_{12}$ (= $[Ba_{12}Nb_{2.4}Ir_{9.6}O_{36}]/3$). The single-crystal X-ray data, especially the high minima and maxima in the difference Fourier maps ($F_o$–$F_c$) close to the Ba atoms, also indicate a certain propensity for stacking faults in this structure, as for many other hexagonal perovskite phases[3]. The above composition is also confirmed by chemical analysis on single crystals which gives an approximate metal composition of $Ba_{4.04}Nb_{0.82}Ir_{3.28}$. Besides, no chlorine was detected in ICP analysis making this flux-method an excellent choice to obtain crystals for studying the intrinsic magnetic properties reliably without contaminating the sample.

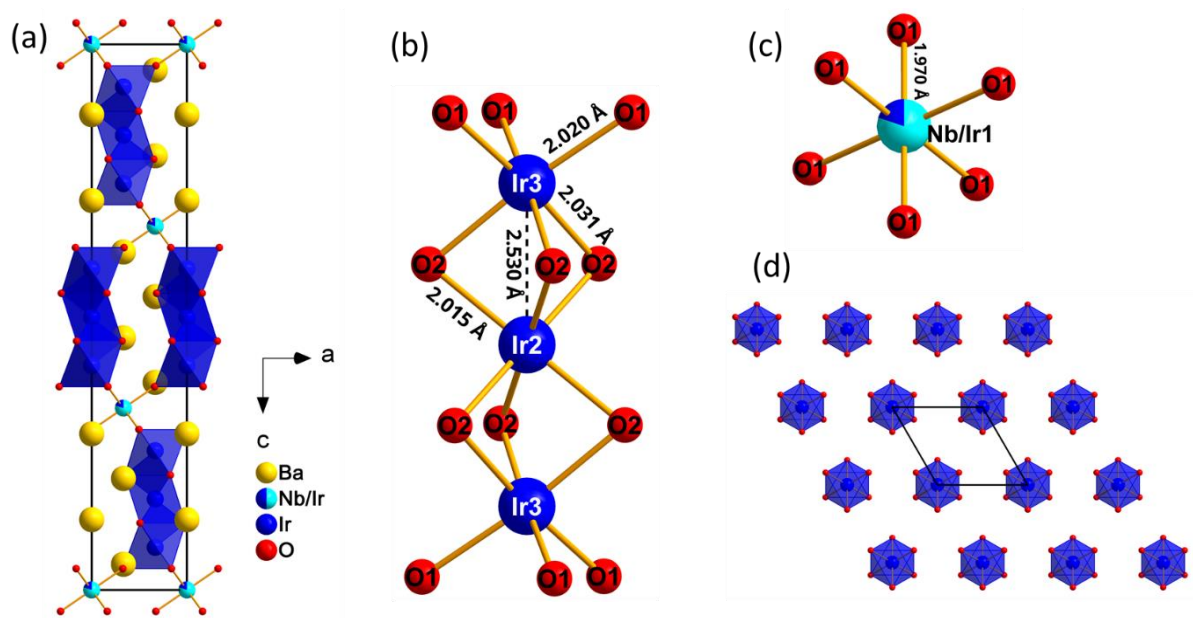

Figure 3. Crystal structure of $Ba_4Nb_{0.8}Ir_{3.2}O_{12}$ (a), coordination around Ir and Nb atoms (b) and (c), and honeycomb-like arrangement of $IrO_6$ octahedra (d).

The compound crystallizes in a 12–R perovskite structure consisting of $Ir_3O_{12}$ trimers of face sharing $IrO_6$ octahedra running along the *c*–axis (figure 3). These trimers of octahedra are arranged in a honeycomb type pattern extended in the *ab*-plane (figure 3c). Three vertices each at the base of the adjacent $Ir_3O_{12}$ trimers are connected by $Nb(Ir)O_6$ octahedra in a corner-sharing fashion resulting in a three-dimensional structure. The Ir–O bond lengths range between 2.01 to 2.035 Å, which matches well with the other known tetravalent iridium-based perovskites.[14-17] The present stoichiometry leads to the presence of three possible valence states of Ir namely, $Ir^{5+}$, $Ir^{4+}$, and $Ir^{3+}$. Although it is difficult to precisely assign a definite valence state to each of the Ir at three different sites, the Bond Valence Sum (BVS) method proved useful[18]. The BVS for Nb/Ir1 comes to 4.74 indicating a pentavalent Ir and Nb. The BVS for corresponding Ir2 and Ir3 atoms was calculated to be 3.978 and 3.819 respectively which implies that the Ir2 is primarily tetravalent and Ir3 is a statistical mixture of 3+ and 4+. A similar charge distribution is also observed in the isostructural ruthenate $Ba_4NbRu_3O_{12}$.[19]

Table 2: Crystallographic parameters of $Ba_4Nb_{0.8}Ir_{3.2}O_{12}$.

| Space group: $R\bar{3}m$ ||||||| 
| Lattice parameters: $a = b = 5.7305(6)$ Å; $c = 28.558(3)$ Å ||||||| 
| Atom | Wyck. | Site symmetry | Occupancy | x/a | y/b | z/c |
| --- | --- | --- | --- | --- | --- | --- |
| Ba1 | 6c | 3m | 1 | 0 | 0 | 0.12896(6) |
| Ba2 | 6c | 3m | 1 | 0 | 0 | 0.28617(5) |
| Nb1/Ir1 | 3a | $-3m$ | 0.80(1)/0.20(1) | 0 | 0 | 0 |
| Ir2 | 3b | $-3m$ | 1 | 0 | 0 | ½ |
| Ir3 | 6c | 3m | 1 | 0 | 0 | 0.41144(3) |
| O1 | 18h | .m | 1 | 0.327(2) | 0.164(1) | −0.0394(4) |
| O2 | 18h | .m | 1 | −0.348(2) | −0.1741(9) | 0.2108(4) |

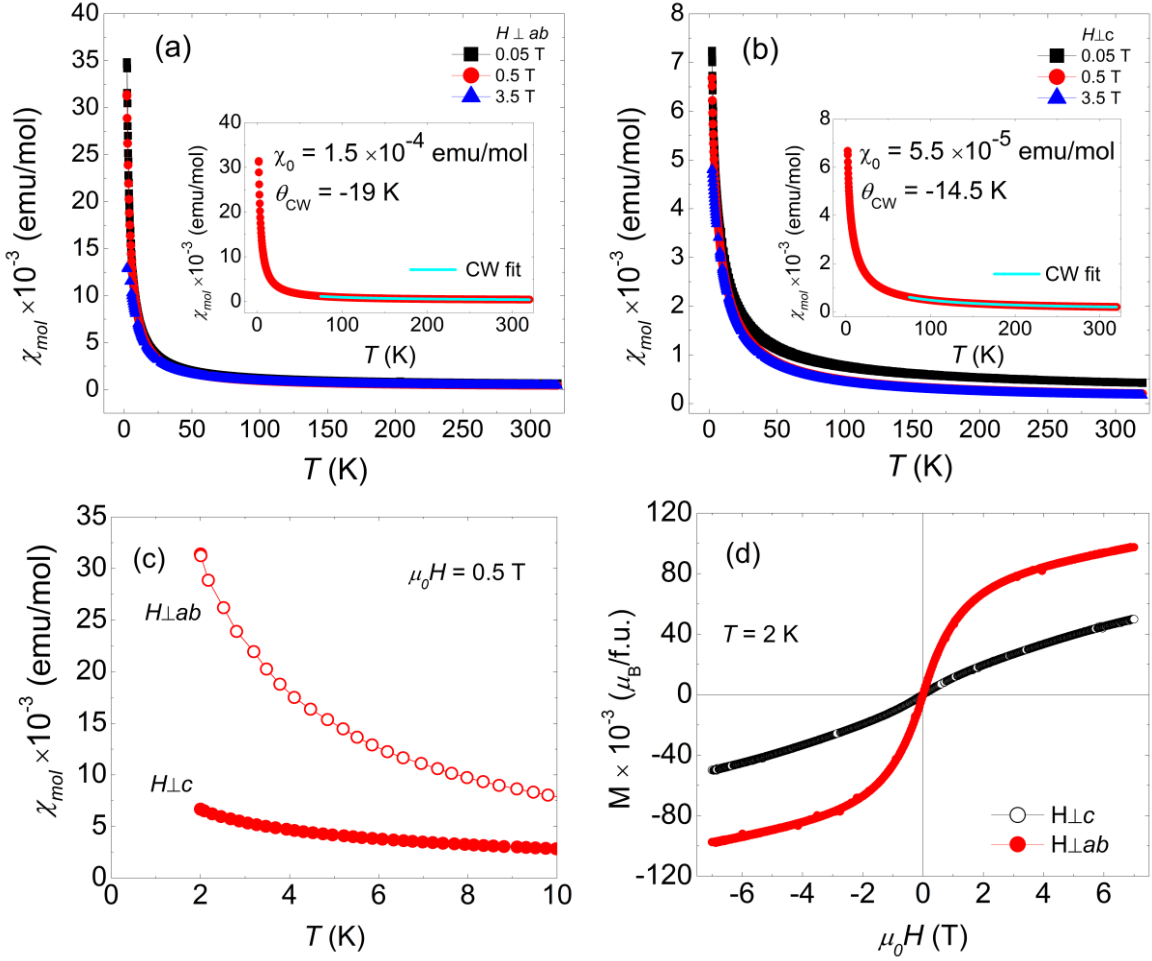

Figure 4. Molar magnetic susceptibility of $Ba_4Nb_{0.8}Ir_{3.2}O_{12}$ in different crystallographic orientations (a) $\mu_0H\perp ab$ (b) $\mu_0H\perp c$, (c) enlarged view of low-temperature susceptibility at $\mu_0H = 0.5$ T and (d) field-dependent magnetization in different crystallographic orientations at 2 K.

Figure 4 shows the temperature dependence of molar magnetic susceptibility ($\chi_m = M/\mu_0H$) of a single-crystal of $Ba_4Nb_{0.8}Ir_{3.2}O_{12}$ in applied magnetic fields from 0.05–3.5 T. The magnetic field was applied in two different directions ($\mu_0H\perp ab$ and $\mu_0H\perp c$). Susceptibility in both directions remains paramagnetic without indication of any long-range magnetic ordering down to the lowest temperature measured, 2 K (Figure 4 c). Insets of figure 4 (a) and (b) show fit of $\chi_m$ to the Curie-Weiss equation; $\chi_m = \chi_0 + C/(T-\theta)$ (where, $\chi_0$ is the temperature independent part of the susceptibility) in the temperature range from 75 K to 320 K, which yields an effective paramagnetic moment of 0.88 and 0.63 $\mu_B$ in $\mu_0H \perp ab$ and $\mu_0H \perp c$ protocols, respectively. The absolute value of susceptibility and the magnetic moments obtained are very close to the results reported earlier on the polycrystalline 'Ba$_4$NbIr$_3$O$_{12}$' sample.[8] Relatively large negative Curie-Weiss temperatures ($\theta_{CW}$) of −19 K and −14.5 K indicate that the interactions between the moments are predominantly antiferromagnetic. The $\chi(T)$ data measured perpendicular to $ab$–plane and $c$ direction show anisotropic behavior, as can clearly be seen in figure 4 (c) the susceptibility is much larger along the $c$ direction. Isostructural compounds Ba$_4$NbM$_3$O$_{12}$ with M = Ru and Rh, however, show magnetic transitions possibly to a spin-glass state at low temperatures.[8,20] The absence of magnetic ordering in Ba$_4$Nb$_{0.8}$Ir$_{3.2}$O$_{12}$ down to the lowest temperature measured (2 K) combined with a relatively large $\Theta_{CW}$, suggests a presence of strong magnetic frustration. A high frustration parameter ($\Theta_{CW}/T_M$) > 9 points toward a possible QSL state. Our observations are very consistent with the results of polycrystalline sample reported by Nguyen et al.[8] However, further low-temperature measurements (below 2 K) are required to eliminate the possibility of any magnetic ordering and thus confirm the QSL state. Figure 4(d) depicts $M$ vs. $H$ curves measured at 2 K with magnetic fields applied both parallel and perpendicular to the $c$ direction. Similar to the $\chi(T)$ data, $M(H)$ data shows anisotropic behavior up to the highest applied magnetic field (7 T) with larger magnetization along the parallel direction. Both M$_{ab}$ and M$_{c-axis}$ curves do not saturate even at a field of 7 T. No hysteretic behavior is observed in both of the $M(H)$ data indicating the absence of any ferromagnetic-like behavior.

To gather more information on the absence of magnetic ordering at low temperatures (down to 2 K) we performed temperature-dependent specific-heat measurements (figure 5). A smooth dependence of specific heat on temperature and the absence of any anomaly eliminates the possibility of magnetic ordering, strengthening the results of susceptibility measurements. The inset of figure 5 shows that there is a tendency of linearity in the $C_p(T)$ data below ~5 K: a similar feature reported by Nguyen *et al* indicating possible spin-liquid phase.[8] Such a feature was found in other QSL candidates as well.[4-6] However, in the present system, measurements down to further lower temperatures would be necessary to investigate this issue. Our results on single crystals agree well with the report of Nguyen *et al* and strongly hints towards a possible QSL state.[8]

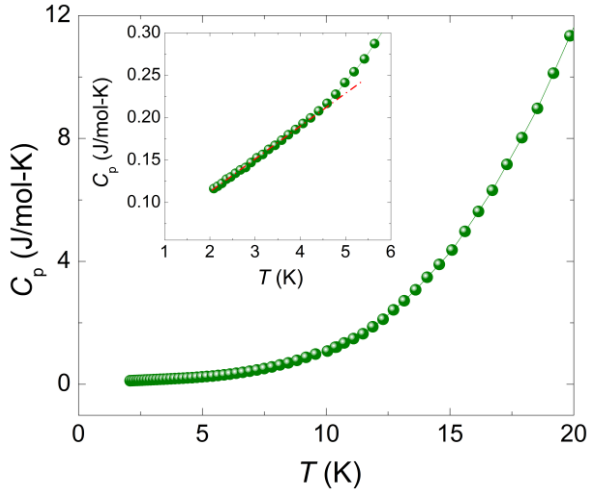

Figure 5: The main panel shows the temperature-dependent specific-heat data. The inset focuses on the low-temperature part of the specific-heat data down to 2 K of a $Ba_4Nb_{0.8}Ir_{3.2}O_{12}$ single crystal. Red line is the guide to the eye.

Figure 6 shows the in−plane resistivity ($\rho_{ab}$) data collected on a single crystal of $Ba_4Nb_{0.8}Ir_{3.2}O_{12}$ in the temperature range of 90−400 K. Resistivity data shows an activation behavior below 200 K typical of a semiconductor. An estimated band gap of 43 meV is evaluated using an Arrhenius fit (log $\rho$ vs 1/T plot) to the temperature-dependent resistivity as shown in the inset of figure 6.

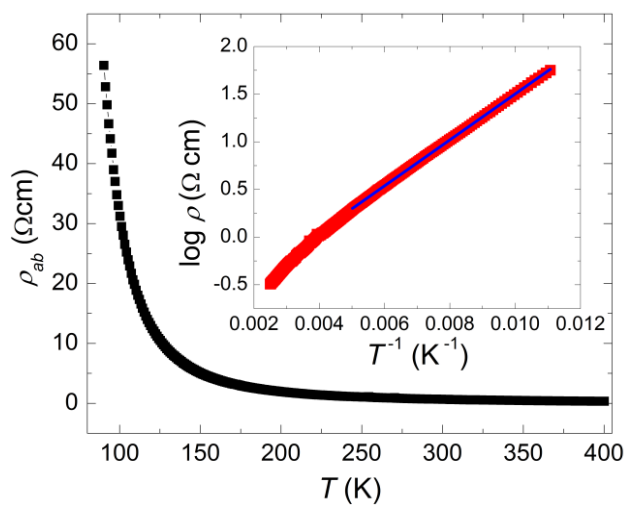

Figure 6: Temperature-dependent resistivity and Arrhenius plot (inset) for $Ba_4Nb_{0.8}Ir_{3.2}O_{12}$ single crystal.

**Conclusion:**

We have successfully grown large single crystals of $Ba_4Nb_{0.8}Ir_{3.2}O_{12}$ using $BaCl_2$ flux. This compound features trimers of face sharing $Ir_3O_{12}$ octahedra and is discussed as a possible QSL candidate. A structure refinement based on single-crystal data confirmed the structure reported previously, including a noticeable mixing of Nb and Ir in one octahedral position. Our magnetic and calorimetric measurements on single crystals confirm the possibility of a spin liquid state existing in this material. The successful growth of large and clean crystals will facilitate a convincing study of this possible exotic state by means of thermodynamic and microscopic probing techniques such as dynamic susceptibility, $\mu$SR and neutron in the mK regime. The availability of high-quality crystals can open new doors to explore the very intriguing and relatively new field of spin liquid oxide materials using a wide range of experimental techniques.

**Supporting Information:** Synthesis scheme, tables for elemental analysis, crystal and refinement data, anisotropic displacement parameters, powder and Laue diffraction patterns and thermal analysis are provided in the supplementary information file. These materials are available free of charge via the Internet at http://pubs.acs.org.

**Author information**

*Email: Gohil.Thakur@cpfs.mpg.de

**Acknowledgments:** The authors thank Walter Schnelle for the specific heat measurements. This work was financially supported by the Deutsche Forschungsgemeinschaft (DFG) within the SFB 1143 "Correlated Magnetism – From Frustration to Topology", project-id 247310070.

**Conflict of interest:** Authors declare no conflict of interest

**Reference:**

1. Shimizu, Y.; Miyagawa, K.; Kanoda, K.; Maesato, M.; Saito, G. Spin liquid state in an organic mott insulator with a triangular lattice. *Phys. Rev. Lett*. **2003**, 91, No. 107001.

2. Nakatsuji, S.; Nambu, Y.; Tonomura, H.; Sakai, O.; Jonas, S.; Broholm, C.; Tsunetsugu, H.; Qiu, Y.; Maeno; Y. Spin disorder on a triangular lattice. *Science* **2005**, 309, 1698-1700.

3. Okamoto, Y.; Nohara, M.; Aruga-Katori, H.; and Takagi, H.; Spin-Liquid State in the S = 1/2 Hyperkagome Antiferromagnet $Na_4Ir_3O_8$. *Phys. Rev. Lett*. **2007**, 99, No. 137207.

4. Mustonen, O.; Vasala, S.; Sadrollahi, E.; Schmidt, K. P.; Baines, C.; Walker, H. C.; Terasaki, I.; Litterst, F. J.; Baggio-Saitovitch, E.; Karppinen, M.; Spin-liquid-like state in a spin-1/2 square-lattice antiferromagnet perovskite induced by $d^{10}$–$d^0$ cation mixing. *Nature Comm*. **2018**, 9, No. 1085.

5. Yamashita, S.; Nakazawa, Y.; Oguni, M.; Oshima, Y.; Nojiri, H.; Shimizu, Y.; Miyagawa, K.; Kanoda, K. Thermodynamic properties of a spin-1/2 spin-liquid state in a κ-type organic salt. *Nat. Phys*. **2008,** 4, 459-462

6. Kelly, Z. A.; Gallagher, M. J.; McQueen, T. M. Electron Doping a Kagome Spin Liquid, *Phys. Rev. X* **2016,** 6, No. 041007.

7. Balz, C.; Lake, B.; Reuther, J.; Luetkens, H.; Schönemann, R.; Herrmannsdörfer, T.; Singh, Y.; Nazmul Islam, A. T. M.; Wheeler, E. M.; Rodriguez-Rivera, J. A.; Guidi, T.; Simeoni, G. G.; Baines C.; Ryll, H. Physical realization of a quantum spin liquid based on a complex frustration mechanism. *Nat. Phys*. **2016**, 12, 942-949.

8. Nguyen L. T.; Cava, R. J. Trimer-based spin liquid candidate $Ba_4NbIr_3O_{12}$. *Phys. Rev. Mater*. **2019**, 3, No. 014412

9. Wilkens, J.; Müller-Buschbaum, H. $Ba_{12}Ir_{12-x}Nb_xO_{36}$ ($x$ = 2.4) - Eine neue verbindung mit 12 R-perowskit-stapelvariante. *J. Alloys and Compd.* **1991,** 176, 141-146.

10. TOPAS-V4.2.0.2: General Profile and Structure Analysis Software for Powder Diffraction Data; Bruker AXS GmbH: Karlsruhe, Germany.

11. Palatinus, L; Chapuis, SUPERFLIP – a computer program for the solution of crystal structures by charge flipping in arbitrary dimensions. G. *J. Appl. Cryst*. **2007**, 40, 786-790.

12. Petricek, V; Dusek, M; Palatinus, L. Crystallographic Computing System JANA2006: General features. *Z. Kristallogr*. **2014**, 229(5), 345-352.

13. Abakumov, A. M; Shpanchenko, R.V; Antipov, E.V; Lebedev, O. I; VanTendeloo, G; Amelinckx, S. Synthesis and Structural Study of Hexagonal Perovskites in the $Ba_5Ta_4O_{15}$–$MZrO_3$ (M = Ba, Sr) System. *J. Solid State Chem*. **1998,** 141, 492—499.

14. Wilkins, J.; Müller-Buschbaum, H.; Zur Kenntnis von $Ba_4Ir_3O_{10}$. *Z. Anorg. Allg. Chemie* **1991,** 592, 79-83,

15. Siegrist, T.; Chamberland, B. L. The crystal structure of $BaIrO_3$. *Journal of the Less-Common Metals* **1991**, 170, 93-99,

16. Powell; A. V. Battle; P. D. Gore; J. G. Structure of $Sr_4IrO_6$ by time-of-flight neutron powder diffraction. *Acta Cryst*. **1993**, C49, 852-854,

17. Longo, J. M.; Kafalas, J. A.; Arnott, J. Structure and properties of the high and low pressure forms of $SrIrO_3$. *J. Solid State Chem*. **1971**, 3, 174-179.

18. Breese, N. E.; O'Keeffe, M. Bond-valence parameters for solids. *Acta Crystallogr*. **1991**, B47, 192-197.

19. Greatrex, R.; Greenwood, N. N. $^{99}$Ru Mössbauer spectroscopic evidence for the presence of discrete ruthenium(III) and ruthenium(IV) ions in the compounds $Ba_4Ru_3NbO_{12}$ and $Ba_4Ru_3TaO_{12}$. *J. Solid State Chem*. **1980**, 31, 281-284.

20. Nguyen, L. T.; Halloran, T.; Xie, W.; Kong, T.; Broholm, C. L.; Cava, R. J.; Geometrically frustrated trimer-based Mott insulator. *Phys. Rev. Mater*. **2018**, 2, No. 054414

# Crystal Growth of Spin-Frustrated $Ba_4Nb_{0.8}Ir_{3.2}O_{12}$: A Possible Spin Liquid Material

*Gohil S. Thakur\*, Sumanta Chattopadhyay, Thomas Doert, T. Herrmannsdörfer and Claudia Felser*

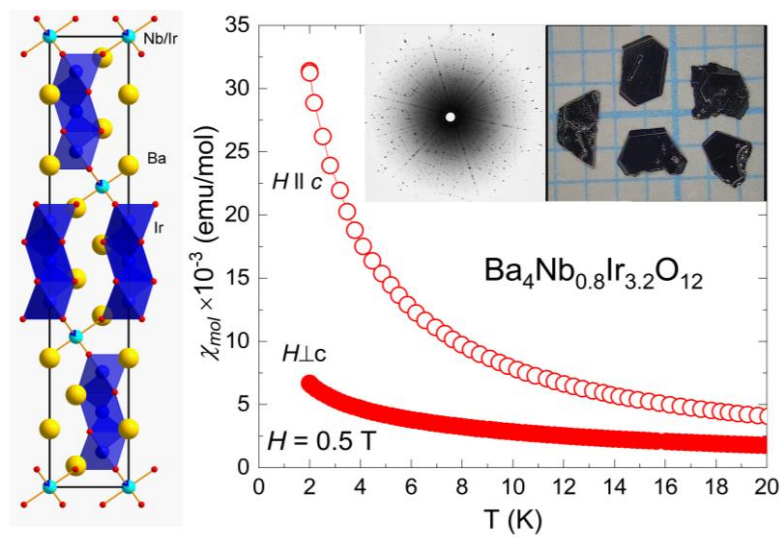

**Synopsis:** Large single crystals of $Ba_4Nb_{0.8}Ir_{3.2}O_{12}$, a possible quantum spin liquid (QSL) oxide material, containing $Ir_3O_{12}$ trimer are grown by an easy and reliable method. This material possesses frustrated spins that do not order magnetically down to 2 K as revealed by magnetic and thermodynamic measurements.

# Supplementary Information

# Crystal Growth of Spin-Frustrated Ba$_4$Nb$_{0.8}$Ir$_{3.2}$O$_{12}$: A Possible Spin Liquid Material

*Gohil S. Thakur [1]\*, Sumanta Chattopadhyay [2], Thomas Doert [3], T. Herrmannsdörfer [2] and Claudia Felser [1]*

[1] Max Planck Institute for Chemical Physics of Solids, 01187 Dresden, Germany

[2] Dresden High Magnetic Field Laboratory (HLD-EMFL), Helmholtz-Zentrum Dresden-Rossendorf, 01328 Dresden, Germany

[3] Faculty of Chemistry and Food Chemistry, Technical University, 01069, Dresden, Germany

**Scheme 1.** Temperature profile for the growth of Ba$_4$Nb$_{0.8}$Ir$_{3.2}$O$_{12}$ single crystals.

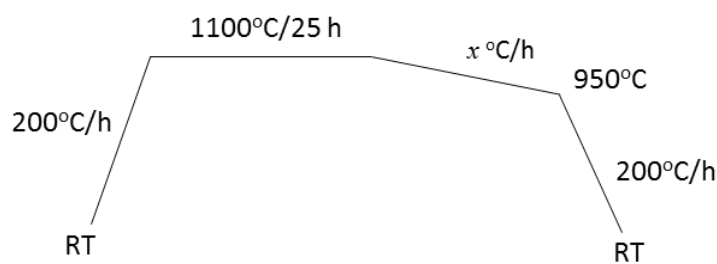

**Table S1.** Metal ratio obtained from SEM-EDX analysis on single crystals.

| S. No. | Element | | |
|---|---|---|---|
| | Ba | Ir | Nb |
| #1 | 29.91 | 23.94 | 6.95 |
| #2 | 31.22 | 23.53 | 7.12 |
| #3 | 32.82 | 24.39 | 7.01 |
| #4 | 32.55 | 23.99 | 7.54 |
| #5 | 32.03 | 22.93 | 7.15 |
| Average | 31.71 | 23.76 | 7.15 |
| Avg. stoichiometry | **Ba$_{4.43}$NbIr$_{3.32}$** | | |

**Table S2**. Crystal and structure refinement data for $Ba_4Nb_{0.8}Ir_{3.2}O_{12}$.

| | |
|---|---|
| Emperical formula | $Ba_4Nb_{0.8}Ir_{3.2}O_{12}$ |
| Temperature | 295 K |
| Wavelength | 0.5638 Å |
| Crystal system | trigonal |
| Space group | $R\bar{3}m$ (148) |
| Unit cell dimensions | $a = b$ = 5.7305(6) Å; $c$ = 28.558(3) Å |
| Volume | 812.2(2) Å$^3$ |
| Z | 2 |
| Density (calculated) | 8.775 Mg/m$^3$ |
| Absorption coefficient | 32.294 mm$^{-1}$ |
| Theta range for data collection | 3.3-25º |
| Index ranges | -8 <=h <= 8, -8 <=k <= 8, -42 <=l <= 42 |
| F(000) | 1798 |
| Reflections used | 5931 |
| Data / restraints / parameters | 620 / 0 / 23 |
| Absorption correction | integration |
| Goodness-of-fit of F$^2$ | 2.36 |
| Final R indices [I > 2sigma(I)] | R1 = 0.0337, wR2 = 0.0773 |
| R indices (all data) | R1 = 0.0525, wR2 = 0.0791 |

**Table S3**. Anisotropic displacement parameters (Å$^2$) for $Ba_4Nb_{0.8}Ir_{3.2}O_{12}$ ($U_{13}=U_{23}=0$)

| Atom | $U_{11}=U_{22}$ | $U_{33}$ | $U_{12}$ |
|---|---|---|---|
| Ba1 | 0.0127(3) | 0.0146(7) | 0.00634(15) |
| Ba2 | 0.0123(3) | 0.0130(7) | 0.00616(14) |
| Nb1 | 0.0091(6) | 0.0114(10) | 0.0046(3) |
| Ir2 | 0.0098(3) | 0.0106(6) | 0.00489(13) |
| Ir3 | 0.0104(2) | 0.0098(4) | 0.00520(11) |

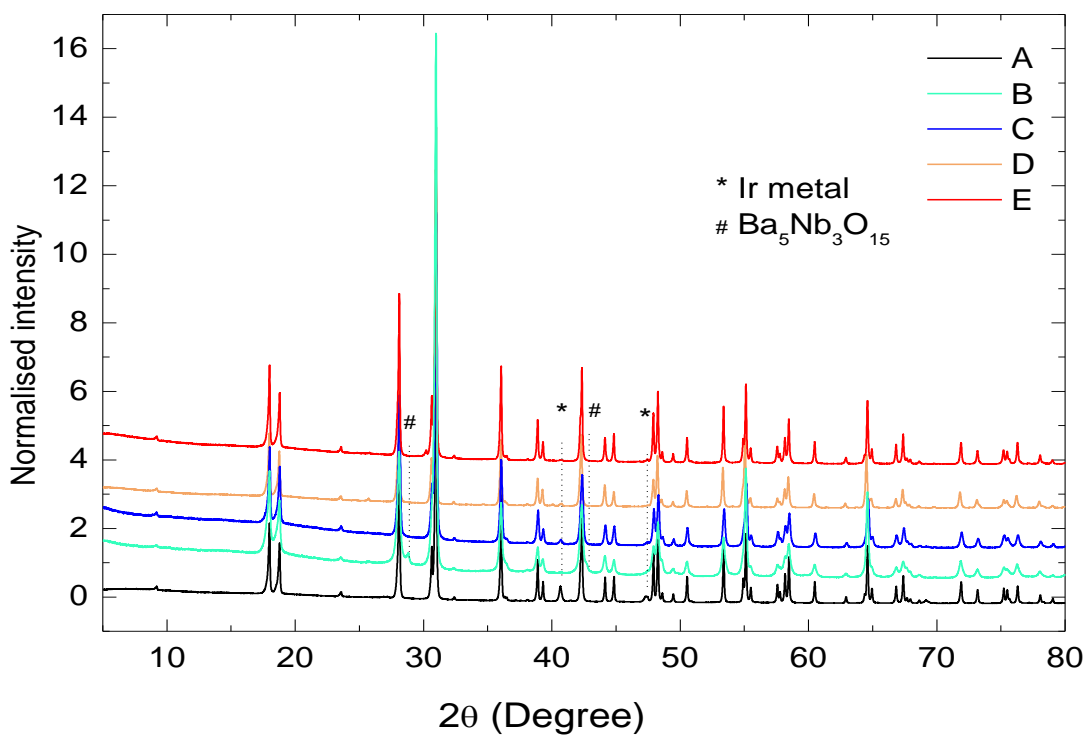

**Figure S1**. Powder diffraction patterns of $Ba_4Nb_{0.8}Ir_{3.2}O_{12}$ samples synthesized at different conditions given in table 1 of the main text.

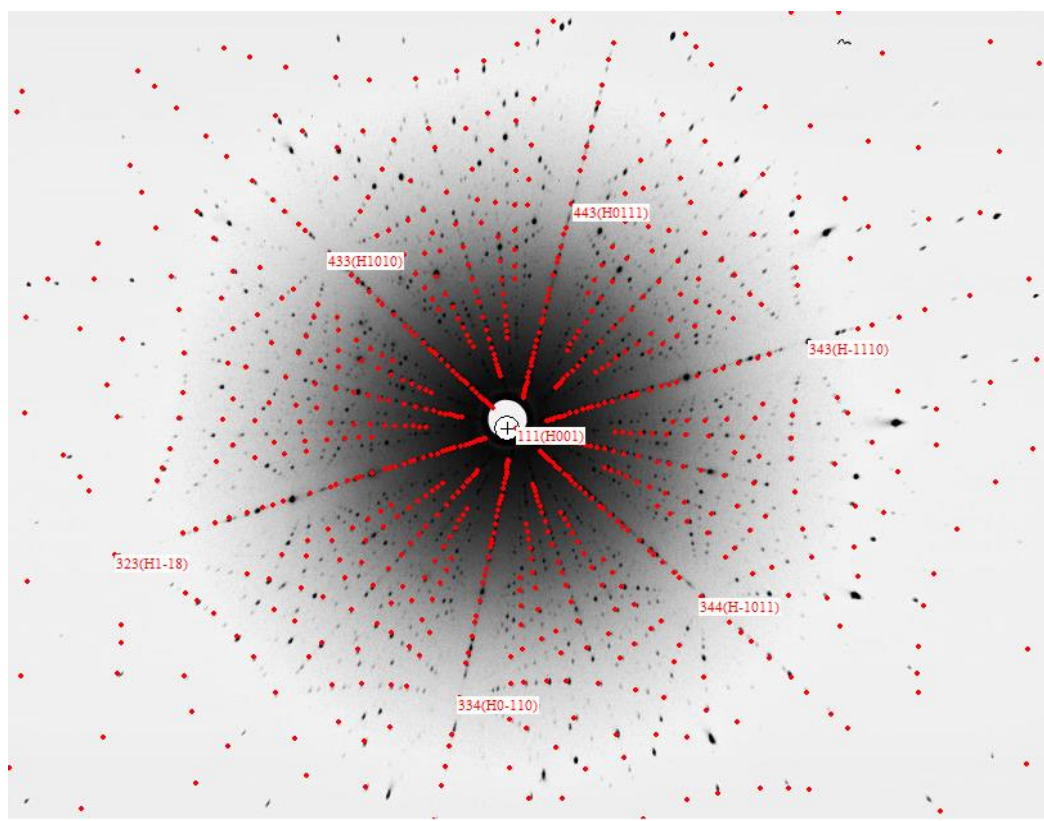

**Figure S2**. Laue diffraction pattern of the flat surface of a $Ba_4Nb_{0.8}Ir_{3.2}O_{12}$ single crystal.

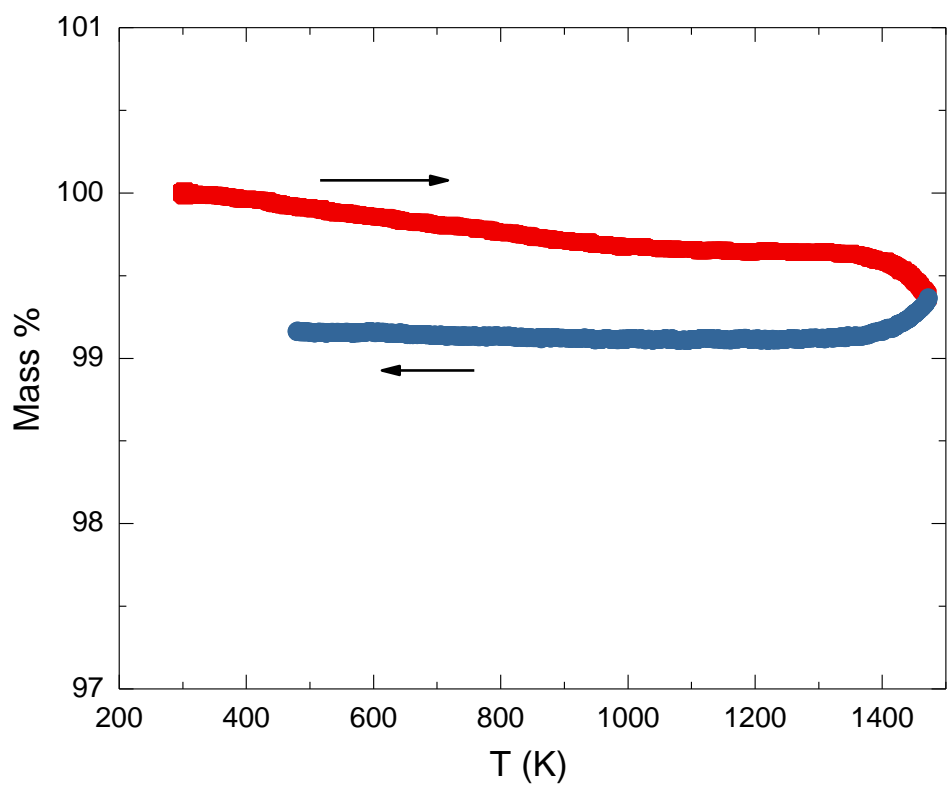

**Figure S3**. Thermal decomposition profile of $Ba_4Nb_{0.8}Ir_{3.2}O_{12}$. The compound does not decompose at least up to 1473 K.